**The Sexy Male Body: Men's Height and Weight are Condition-Dependent, Sexually Selected Traits**

**David Giofrè,[1] David C. Geary,[2] & Lewis Halsey[3]**

[1]University of Genoa, Genoa (Italy)

[2]University of Missouri, Columbia (USA)

[3]University of Roehampton, London (UK)

**Author Contact:**

**David Giofrè**

**david.giofre@unige.it**

**Declaration of interests**

**The authors declare no competing interests.**

**Footnotes**

**Electronic supplementary material is available online at https://doi.org/10.6084/m9.figshare.c.7622489.**

**Abstract**

On average men are taller and more muscular than women, which confers on them advantages related to female choice and during physical competition with other men. Sexual size dimorphisms such as these come with vulnerabilities due to higher maintenance and developmental costs for the sex with the larger trait. These costs are in keeping with evolutionary theory that posits large, elaborate, sexually selected traits are signals of health and vitality because stressor exposure (e.g. early disease) will compromise them (e.g. shorter stature) more than other traits. We provide a large-scale test of this hypothesis for the human male and show that with cross-national and cross-generational improvements in living conditions, where environmental stressors recede, men's gains in height and weight are more than double those of women's, increasing sexual size dimorphism. Our study combines evolutionary biology with measures of human wellbeing, providing novel insights into how socio-ecological factors and sexual selection shape key physical traits.

# 1. Introduction

Sexual size dimorphism (SSD) is observed in many species and can result from the sexes occupying different feeding niches, larger females being more fecund or larger males having advantages related to female choice or male–male competition [1,2]. The latter is common in terrestrial mammals, including humans [3–5]. Men are on average taller and heavier than women across nations, which likely reflects an evolutionary history of physical male–male competition for status and resource control [6]. Consequently, this is why men assess the formidability of potential opponents using physical cues, such as upper body strength [7], and women find taller, muscular men with a relatively large overall body mass (but not obese) particularly attractive [8].

Larger size, particularly when coupled with, for example, relatively greater muscle mass [9], comes with higher developmental and maintenance costs. Full development of these traits is particularly dependent upon long-term physical and physiological health [10], making their expression dependent on body condition [11]. This is not the case for all sexually dimorphic traits, such as face structure, but might be the case for many traits showing sex differences, including those favouring women (e.g. episodic memory) and related to female–female competition [3,12,13]. The SSD for height is large and thus should provide a particularly sensitive index of sex differences in vulnerability to early environmental conditions. In other words, men's height should be especially sensitive to disruption by infectious disease and nutritional shortfalls during development; the associated stunted growth is predictive of health and longevity in adulthood [14,15]. In contrast, women's smaller stature and smaller metabolically demanding organs make them less susceptible to environmental fluctuations in available nutrients and calories [16].

Consequently, we hypothesize that women's physical size will be less compromised by early environmental stressors than men's size, with men being disproportionately smaller when they have grown up in poor environments. In turn, the degree of SSD in a stress-free population is greater than in a stressed population. Moreover, in stress-free environments, people are relatively uncompromised in their growth, while in stressful ones, some people are more compromised than others, and thus we hypothesize that variation in size between individuals is greater in more stressful scenarios. If men's growth is more susceptible to developmental perturbations, then they should be more variable in height than women in stressful environments and less variable in stress-free environments.

We tested these hypotheses using a multinational survey of sex differences in height and weight conducted by the World Health Organization (WHO) [15] and a measure of national levels of human well-being—the human development index (HDI). Individuals living in countries with low HDI scores are more likely to suffer from infectious diseases, chronic poor nutrition and larger overall disease burden than are individuals living in countries with high HDI scores [17]. We predicted men will be taller, weigh more and be generally less variable in these traits in countries with high HDI scores. As a robustness check, we conducted the same analyses on a vetted dataset on adult height from Wikipedia. To assess within-country changes in height during a period when living conditions were improving, we included a cross-sectional study of people born in the UK between 1900 and 1958 [18]. Finally, we tested the same hypotheses with an alternative to the HDI—the World Bank's Gini index of income inequality [19]. If men's heights and weights are especially sensitive to developmental conditions, higher inequality should be more strongly related to men's than women's heights and weights.

## 2. Materials

The first data source we analysed was self-reported height and weight provided in 2003 by the WHO for 69 countries. Inspection of the data indicated that a small number of data points were spurious. To deal with this, we excluded individuals with a reported height of less than 1 m or a weight lower than 25 kg. We also excluded individuals with a body mass index (BMI) higher than 30 (i.e. obese status) in our main analyses; analyses including these individuals revealed the same patterns (electronic supplementary material, Section 1). Countries with fewer than 100 participants for either males or females were also excluded because small samples can skew estimates of physical dimorphisms [20,21]. Excluded countries were Bangladesh, the Dominican Republic, Guatemala, Mali, Myanmar, Nepal and Pakistan. Overall, 14% of the original sample was excluded. The main analyses included 135 645 participants (62 681 males) across 62 countries, with samples ranging from 490 (Slovenia) to 19 750 (Mexico). We then reran these analyses to include the aforementioned countries with small samples and not exclude any participants, and performed some meta-analytic analyses, which confirmed the patterns in the main analyses (electronic supplementary material, Section 1).

The second data source was adult height data from Wikipedia [22]. The sources cited therein reveal that most data come from peer-reviewed publications or official health reports (e.g.

from WHO), and largely after 1999. These data were independent from WHO 2003. There are a few less reliable sources (e.g. newspaper stories) and these were deleted from our dataset, resulting in the removal of the data associated with Ecuador, Israel, Russia and Kenya. Separate data were included for males ($n$ = 130 countries) and females ($n$ = 147).

The third source was Kuh *et al.*'s [18] compilation of adult heights from several UK studies (e.g. National Survey of Health and Development). The sample included 49 180 adults (22 441 males), aged 23−26 years, and was categorized into ten 5-year date of birth intervals, from 1905 to 1958. The mean sample size per sex and date of birth cohort was 2459 and ranged from 117 (females born in 1905 or earlier) to 5571 (females born in 1958). The standard deviations (s.d.) for two date of birth cohorts were outliers with values ranging from 2.93 to 3.71 s.d. higher or lower than adjacent values (males 1910−1915, females 1925−1930), and these were replaced by the mean of the two adjacent values. The same results were found with the exclusion of these two cohorts (electronic supplementary material, Section 2).

Historical data for the HDI were obtained for 2003 (https://hdr.undp.org/data-center/human-development-index#/indicies/HDI). The three components of the HDI are life expectancy, years of schooling and *per capita* income, which in combination provide a summary index of average levels of well-being for citizens within each participating country. The Gini index was available for 58 countries between 2000 and 2006 (https://data.worldbank.org/indicator/SI.POV.GINI).

**(a) Data and code availability**

The data analysed in this study were obtained from existing sources. WHO data are restricted and available upon request from the original data holder, but other data are included in the OSF online repository (https://osf.io/j2h4w/?view_only=4e282674d4154b5fbeedb3eca09f6282).

**(b) Quantification and statistical analysis**

R, v. 4.3.2, was used for the analyses (electronic supplementary material). For the WHO data, descriptive statistics for each country were calculated from the height and weight of individuals to obtain means and s.d. [23]. Sexual dimorphisms in height and weight were calculated as the male/female ratio of mean values, subsequently $\log_{10}$-transformed, as is recommended for ratio data [24]. Sex, HDI and their interaction were regressed on height and weight. The Breusch–Pagan (BP) test was conducted to assess heteroscedasticity using the 'lmtest' package. To further ensure robustness, analyses were repeated using robust standard errors (via vcovHC in the 'sandwich' package).

We repeated the analyses using the adjusted mean height and weight variables, standardized relative to the UK's (GBR) mean values for both males and females. This adjustment allowed us to compare proportional differences across countries relative to a consistent reference point, which revealed the same pattern (electronic supplementary material, Section 1).

For the average human height by country data gleaned from Wikipedia [22], results were checked for their veracity and four countries were not included in analyses (electronic supplementary material, Section 3). We compared the two beta coefficients using the method outlined by Cohen *et al.* (electronic supplementary material, Section 3). The two regression beta coefficients were compared using Cohen's formula [25]—calculating the difference between the coefficients determining the standard error of this difference and then computing a Z-score to assess whether the observed difference was statistically significant. These analyses were also performed on a reduced sample size ($n = 128$) where data for every country were available for both males and females, which produced very similar results (not reported).

For the UK data, height and s.d. were regressed against the date of birth cohort, sex and interaction. Sexual dimorphism in height was regressed against date of birth cohort. The analyses were also conducted using proportional changes in height, which revealed the same pattern (electronic supplementary material, Section 2).

## 3. Results

For the WHO height data, the model with the interaction between sex and HDI was significant and explained considerable variance (figure 1*a*, $F_{3,120} = 3.53$, $p = 0.001$, $R^2 = 0.718$). The interaction was significant, $F_{1,120} = 10.87$, $p < 0.001$, as were sex, $F_{1,120} = 241.40$, $p < 0.001$, and HDI, $F_{1,120} = 63.95$, $p < 0.001$: each 0.2 increase in HDI was associated with an average height increase of approximately 1.68 cm for females and 4.03 cm for males. Height sexual dimorphism was linearly correlated with HDI ($r = 0.80$, $p < 0.001$; figure 1*c*). The test results were not statistically significant for the BP test, BP(3) = 3.368, $p = 0.338$, indicating no evidence of heteroscedasticity. The results using robust standard errors indicated the interaction term remained significant ($t = 3.343$, $p = 0.001$).

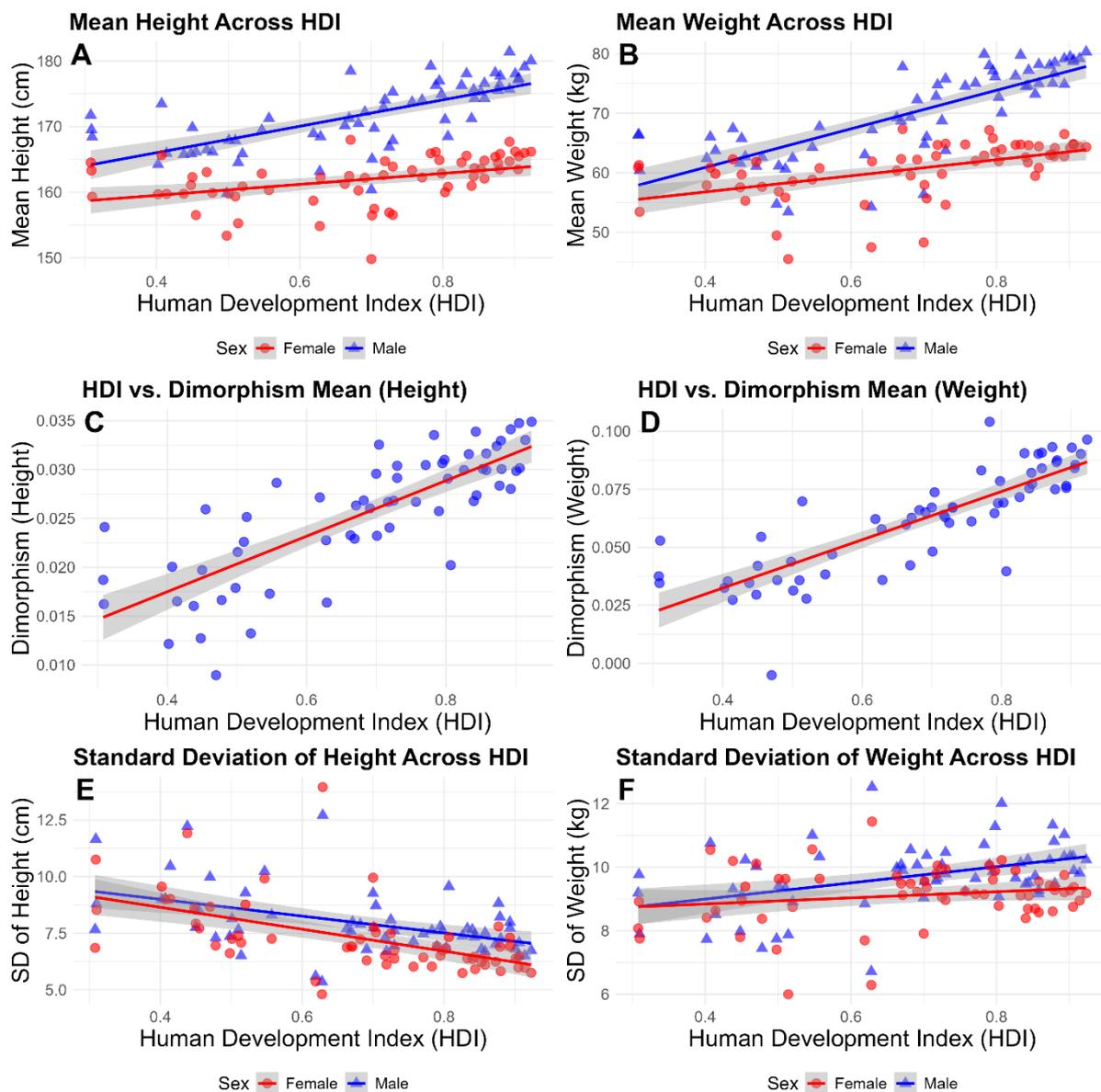

*Figure 1.* Panels A and B respectively show the positive relationships between HDI and median height and weight for males (blue) and females (red). Panels C and D respectively show that sexual dimorphisms in height and weight increase with HDI. Panels E and F respectively show the relationships between inter-individual variation in height and weight and HDI. Each data point represents a unique country. Linear fits are presented with 95% confidence intervals.

For weight, the model with the interaction between sex and HDI was significant and explained considerable variance (figure 1*b*; $F_{3,120} = 4.39$, $p = 0.001$, $R^2 = 0.692$). The interaction between sex and HDI was significant, $F_{1,120} = 18.13$, $p < 0.001$, as were sex, $F_{1,120} = 144.64$, $p < 0.001$, and HDI, $F_{1,120} = 107.34$, $p < 0.001$: each 0.2 increase in HDI was associated with an average weight increase of 2.70 kg for females and 6.48 kg for males. Weight sexual dimorphism was linearly correlated with HDI ($r = 0.83$, $p < 0.001$; figure 1*d*). The BP test was not significant, BP(3) = 6.084, $p = 0.108$, indicating no evidence of heteroscedasticity. The interaction term remained significant ($t = 4.590$, $p < 0.001$) when using robust standard errors.

Variation in height (s.d.) was negatively linearly related to HDI (average correlation for males and females, $r = -0.51$, $p < 0.001$) (figure 1*e*), whereas variation in weight was positively linearly related to HDI (average correlation for males and females, $r = 0.31$, $p < 0.001$) (figure 1*f*). The interaction between sex and HDI was not significant for either height ($p = 0.386$) or weight ($p = 0.096$). Therefore, we refitted each model including only the effect of HDI and sex.

Analysis of the average human height by country data gleaned from Wikipedia confirmed the WHO data patterns (figure 2; electronic supplementary material, Section 3).

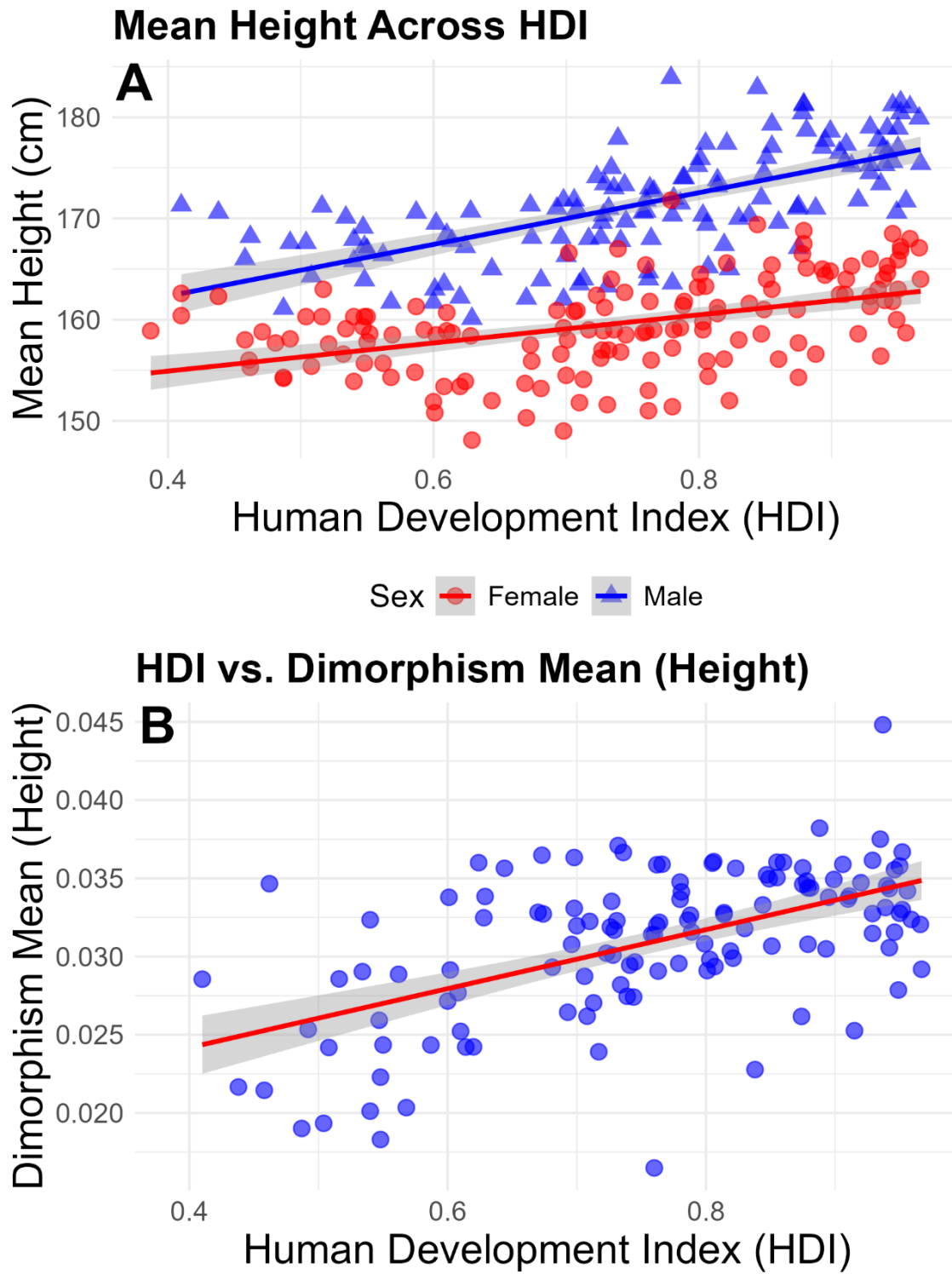

Figure 2. (*a*) The positive relationships between HDI and mean height for males (blue) and females (red). (*b*) Sexual dimorphisms in height increase with HDI.

There was a significant sex by date of birth cohort effect for the UK height data ($p < 0.001$). Female height increased by 0.25 cm for every 5-year interval ($R^2 = 0.72$, $p = 0.002$), and male height increased by 0.69 cm ($R^2 = 0.95$, $p < 0.001$) (figure 3*a*). There was a 0.04 cm reduction in female s.d. across date of birth cohorts, but this was not significant ($R^2 = 0.27$, $p = 0.122$), as compared to a significant 0.23 cm reduction for males ($R^2 = 0.92$, $p < 0.0001$) (figure 3*b*). The interaction for s.d. was also significant ($p < 0.001$), confirming men's s.d. declined more steeply than women's s.d. across cohorts. Height sexual dimorphism was significantly correlated with date of birth cohort ($R^2 = 0.95$, $p < 0.001$).

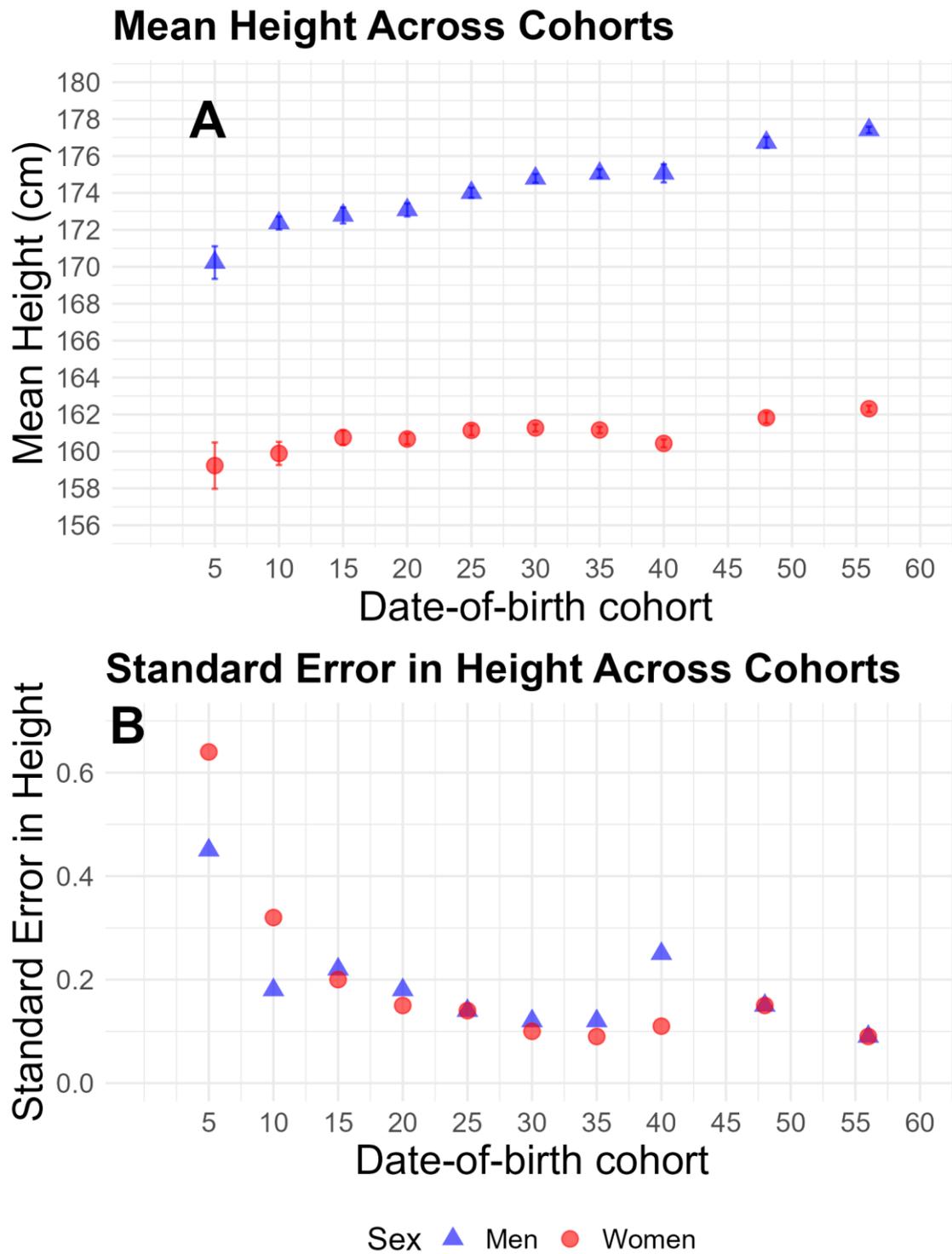

Figure 3. (*a*) Changes in height and (*b*) s.e. in height for date of birth cohorts born before 1905 (5 on *x*-axis) and 1958. Error bars represent 95% CIs. Note that in both panels the *y*-axis does not start at zero.

Finally, regressions between height and the Gini coefficient (electronic supplementary material, Section 1) for the WHO data revealed a significant interaction between sex and Gini, $F_{3,112} = 73.92$, $p < 0.001$, $R^2 = 0.664$. Each unit increase in Gini (i.e. higher inequality) was associated with an average height decrease of approximately 0.14 cm for females and 0.31 for males. Height SSD decreased as inequality increased ($r = -0.58$). The interaction between sex and Gini was also significant for weight, $F_{3,112} = 35.67$, $p < 0.001$, $R^2 = 0.489$. Each unit increase in Gini was associated with an average weight decrease of approximately 0.13 kg for females and 0.39 kg for males. Again, SSD for weight decreased as inequality increased ($r = -0.62$).

## 4. Discussion

Our cross-national analyses suggest that as the social and ecological conditions of nations improve, including reductions in overall disease burden [17], people's height and weight increase, but more than twice as much in men as in women resulting in greater SSD. It is possible that this sex difference is confounded by population differences in maximum potential height under optimal living conditions, and thus our estimates of the rate of height increase with increasing HDI may be biased upward. However, any such bias is not likely to be substantive, given that the same pattern emerged in our within-nation analysis—in the UK, men's height gains were 2.76 times larger than women's in the first half of the twentieth century, where over 90% of the population at that time was White European [26]. Other studies confirm substantive within-nation effects: Nigerian men who have grown up in nutritionally stressed regions are 7.5 cm shorter than their better-nourished peers, whereas women are just 3.2 cm shorter [27].

As predicted, in our across-country analyses inter-individual variation in height was lower in countries with better living conditions, but there was not our hypothesized difference in this relationship between men and women. The within-country (UK) pattern, however, yielded our hypothesized sex differences, with larger decreases in variation for men than women as living conditions improved. The reasons for the different cross-country and within-country effects are not yet clear but might be explained by greater noise in the across-country data due to variation between nations and will require follow-up studies. Variability in weight was actually greater in nations with higher HDIs, and this might be explained by richer countries having access to more energy-dense processed foods enabling considerable weight gain by select individuals, although we partially controlled for this by excluding individuals who would be considered obese (BMI > 30).

Based on our main finding of SSD being greater in more favourable environments, it is clear that the development and maintenance of gross morphology is more sensitive to living conditions in men than in women, at least in terms of height and weight. Similarly, for example, to the horns of the ibex [28], men's height and weight fluctuate with developmental conditions in ways consistent with a sexually selected condition-dependent trait [3,11]. More practically, our results confirm Tanner's [10] proposal that height is a useful biomarker for assessing population health. We add to this by showing that men's height and the SSD in height are particularly sensitive to early conditions and may be especially useful biomarkers for tracking population changes in health and for assessing population-level SSDs for other traits, including those that typically favour women [3,11]. At the same time, our results need to be interpreted with caution given their correlational nature and followed up with longitudinal studies that track cohorts varying in early exposure to stressors.

**Electronic Supporting Material**

**The Sexy and Formidable Male Body: Men's Height and Weight are Condition-Dependent, Sexually Selected Traits**

David Giofrè[1], David C. Geary[2] &
Lewis G. Halsey[3]

**Statistical analyses**

All Analyses were performed using R, version4.3.2 (R Core Team, 2023). Ratios often exhibit right-skewed distributions, which can violate the normality assumptions of many statistical analyses. Therefore, we decided to apply a logarithmic transformation to these ratios, which can mitigate skewness, stabilize variance, and make the data more suitable for statistical procedures. This approach is commonly recommended in statistical literature (Emerson, 2014).

**Section 1. WHO analyses**

**Analyses on a larger sample of countries and subjects**

We tried to replicate our results using as many countries as possible and using all available data. Some countries did not have any available data for one sex or both, making it impossible to perform any analysis, Dominican Republic, Myanmar, Nepal. One country had a very limited sample size, Bangladesh, with 4 males and 6 females, while another Mali, presented inconsistency with the data. The latter was probably caused by the simultaneous use of the cm and inches in the data, which resulted in an overall height for males of 115 cm for males and 95 for females, which is clearly implausible. Therefore, these countries were removed from the analyses. At the same time, we did not include any filter including subjects with height lower than 1m and with a BMI over 30. Overall, this resulted in a larger sample 166,571 (45.77% males).

Analyses are reported in (Figure S01).

For height, the linear model including the interaction between sex and HDI was statistically significant and explained a large portion of the variance, $F(3, 124) = 72.61$, $p <.001$, $R^2=.629$. The interaction between sex and HDI was significant, $F(1, 124)=4.54$, $p=.0351$, as were the main effects of sex, $F(1, 124) = 136.10$, p<.001, and HDI, $F(1, 124)=77.19$, $p<.001$. Sexual dimorphism was positively correlated with HDI, $r=.64$, $p<.001$. The Breusch-Pagan test indicates no evidence of heteroscedasticity, $BP(3)=7.78$, $p=.051$. Robust standard errors confirm the significance of the interaction term, $t=2.049$, $p=.042$.

For weight, the linear model including the interaction between sex and HDI was statistically significant and explained a large portion of the variance, $F(3, 124) = 56.67$, $p<.001$, $R^2=.567$. The interaction between sex and HDI was significant, $F(1, 124)=10.04$, $p=.002$, as were the main effects of sex, F(1, 124) = 78.07, p < 0.001, and HDI, F(1, 124) = 81.91, p < 0.001. Sexual dimorphism was positively correlated with HDI, $r=.80$, $p<.001$. The Breusch-Pagan test indicates no evidence of heteroscedasticity, $BP(3)=6.282$, $p=.099$. Robust standard errors confirm the significance of the interaction term, t=3.81, p<.001.

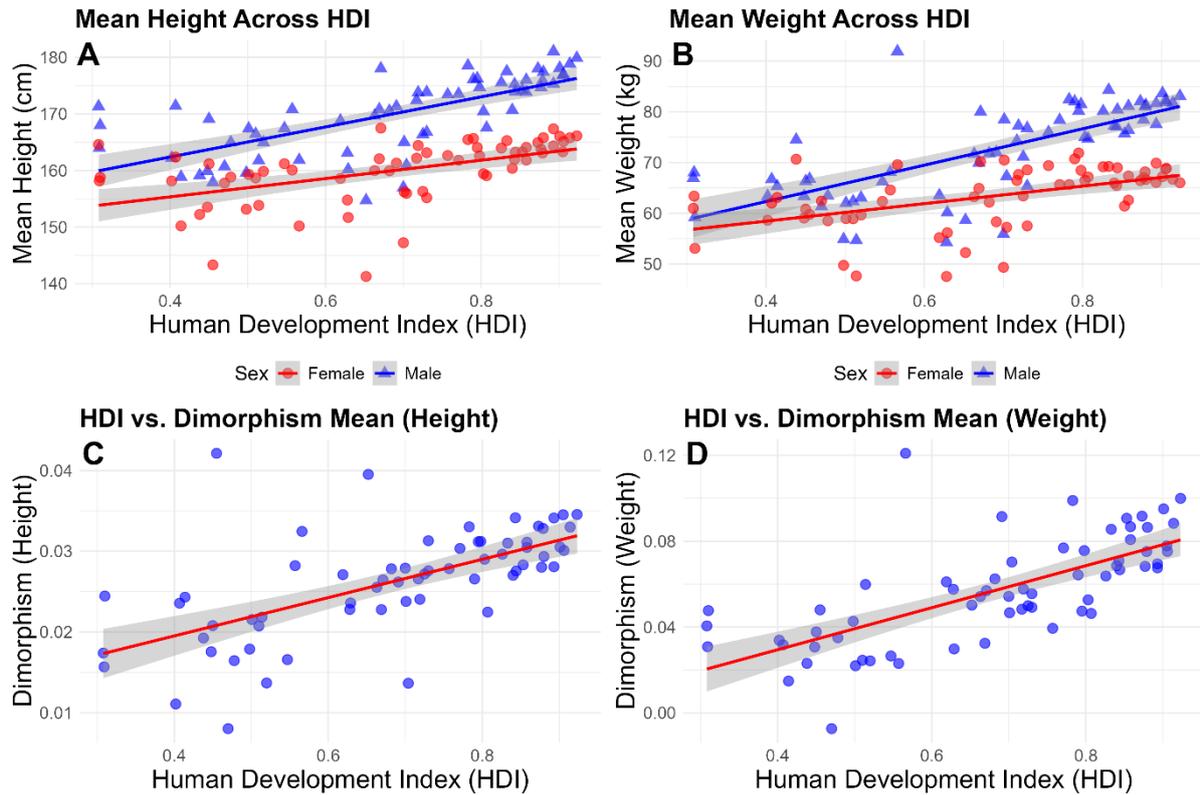

*Figure S01.* Panels A and B, respectively, show the relationships between HDI and mean height and weight for males (blue) and females (red). Panels C and D, respectively, show that sexual dimorphism in height and weight increases with Gini. Each data point represents a unique country. Linear fits are presented with 95% confidence intervals.

**Meta-analyses**

We conducted the meta-regressions using the metafor package in R (Viechtbauer, 2010), a widely used tool for conducting meta-analytical models. Specifically, we employed the rma() function with a restricted maximum likelihood (REML) estimator to fit mixed-effects models. The effect sizes (log-transformed sexual dimorphism values for height and weight) were entered as the dependent variables (yi), with their respective standard errors (sei) included to account for precision in the estimates. The Human Development Index (HDI) was included as a moderator (mods) to assess its relationship with sexual dimorphism. This approach accounts for variability between countries by weighting each data point based on its precision, while also modelling residual heterogeneity to capture unexplained variability. Results across 64 countries are presented in Figure S02 (Panel A height, Panel B weight). The results for height indicated a significant positive effect of HDI on sexual dimorphism ($\beta$=0.0256, *SE*=0.0030, *p*<.0001, *95% CIs* [0.0197, 0.0314]), suggesting that higher HDI levels are associated with greater sexual dimorphism in height. The model accounted for 59.02% of the heterogeneity, though substantial residual heterogeneity remained ($I^2$ = 94.21%). The results for weight showed a significant positive effect of HDI on sexual dimorphism ($\beta$=0.1018, SE=0.0108, *p* <.0001, *95% CIs* [0.0806, 0.1230]), indicating that higher HDI levels are associated with greater sexual dimorphism in weight. The model explained 60.43% of the heterogeneity, but substantial residual heterogeneity remained ($I^2$=95.56%). Results were replicated using a bootstrap approach and were nearly identical, with coefficients and confidence intervals as follows: height, $\beta$=0.025, 95% *CIs* [0.019, 0.032], and weight, $\beta$=0.102, *95% CIs* [0.081, 0.127].

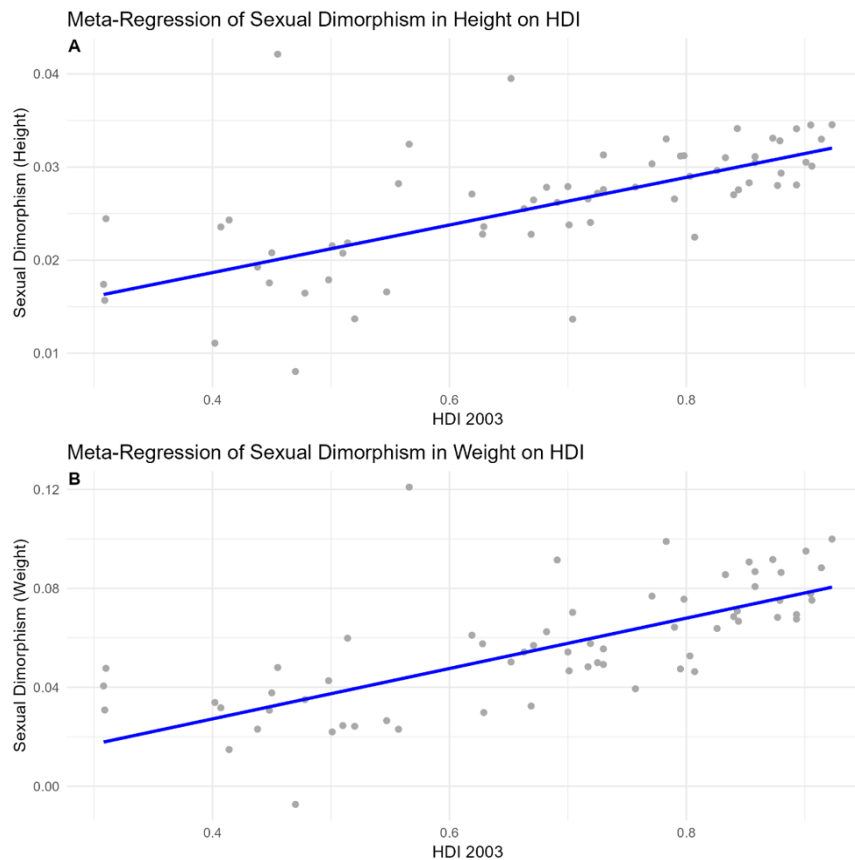

*Figure S02.* Meta-regression of Dimorphism using HDI as a moderator.

## Proportional Effects for WHO

We repeated the analyses using the adjusted mean height and weight variables, standardized relative to the United Kingdom's (GBR) mean values for both males and females. This adjustment allowed us to compare proportional differences across countries relative to a consistent reference point (Figure S03).

For adjusted mean height, the linear model including the interaction between sex and HDI was statistically significant and explained a large portion of the variance, $F(3, 120)=30.33$, $p<.001$, $R^2=.431$. The interaction between sex and HDI was significant, $F(1, 120)=9.02$, $p=.003$, as were the main effects of sex, $F(1, 120)=19.61$, $p<0.001$, and HDI, $F(1, 120)=62.37$, $p<0.001$. Each 0.1 increase in HDI is associated with an average proportional height increase of approximately 0.0051 (0.51% of GBR mean height) for females, and 1.14% for males. Adjusted height sexual dimorphism was positively correlated with HDI, $r=0.80$, $p<.001$. The Breusch-Pagan test indicates no evidence of heteroscedasticity, $BP(3)=2.905$, $p=0.407$. Robust standard errors confirm the significance of the interaction term, $t=3.059$, $p=.003$.

For adjusted mean weight, the linear model including the interaction between sex and HDI was statistically significant and explained a large portion of the variance, $F(3, 120)=43.81$, $p<.001$, $R^2=.511$. The interaction between sex and HDI was significant, $F(1, 120)=10.60$, $p=.001$, as were the main effects of sex, $F(1, 120)=20.38$, $p<.001$, and HDI, $F(1, 120)=100.46$, $p<.001$. Each 0.1 increase in HDI was associated with an average proportional weight increase of approximately 0.0215 (2.15% of GBR mean weight) for females, and 4.22% for males. Adjusted weight sexual dimorphism was positively correlated with HDI, $r=0.80$, $p<.001$. The Breusch-Pagan test indicates no evidence of heteroscedasticity, $BP(3)=5.526$, $p=.137$. Robust standard errors confirm the significance of the interaction term, $t=3.540$, $p<.001$.

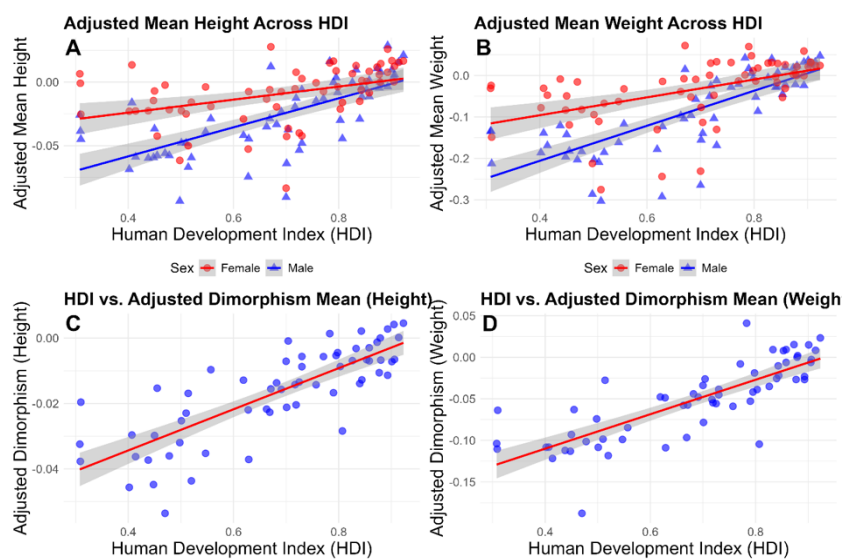

*Figure S03.* Panels A and B, respectively, show the positive relationships between HDI and mean height and weight for males (blue) and females (red), using GBR as the referencing variable. Panels C and D, respectively, show that sexual dimorphism in height and weight increases with HDI. Linear fits are presented with 95% confidence intervals.

**Gini analyses**

All the analyses using the WHO 2003 dataset were repeated using Gini data that were available for 58 countries. To obtain as possible we averaged Gini data from 2000 to 2006. The results were like those obtained using HDI (Figure S04). In fact, HDI and Gini, as expected, were negatively related ($r=-.47$).

The linear model including the interaction between sex and the Gini coefficient was statistically significant and explained a large portion of the variance, $F(3, 112)=73.92$, $p<.001$, $R^2=.664$. The interaction between sex and Gini was significant, $F(1, 112)=4.54$, $p=.035$, as were the main effects of sex, $F(1, 112)=183.84$, $p<.001$, and Gini, $F(1, 112)=33.37$, $p<.001$. Each unit increase in Gini was associated with an average height decrease of approximately 0.14 cm for females, and 0.31 cm for males. Height sexual dimorphism was negatively correlated with GINI ($r=-.58$). The Breusch-Pagan test was not significant, $BP(3)=2.535$, $p=.469$, which indicates no evidence of heteroscedasticity. To ensure robustness, analyses were repeated using robust standard errors via the sandwich() R-package. The interaction term remains statistically significant, $t=-2.444$, $p=.016$.

For weight, the linear model including the interaction between sex and the Gini coefficient was statistically significant and explained a large portion of the variance, $F(3, 112)=43.81$, $p<.001$, $R^2=.489$. The interaction between sex and Gini was significant, $F(1, 112)=5.33$, $p=.023$, as were main effects of sex, $F(1, 112)=80.93$, $p<.001$, and Gini, $F(1, 112)=20.76$, $p<.001$. Each unit increase in Gini was associated with an average weight decrease of approximately 0.13 kg for females and 0.39 kg for males. Weight sexual dimorphism is negatively correlated with Gini ($r=-.62$). The Breusch-Pagan test again indicates no evidence of heteroscedasticity, $BP(3)=5.656$, $p=.130$. Robust standard errors confirm the significance of the interaction term, $t=-3.012$, $p=.003$.

Overall, these results suggest that higher income inequality as measured by the Gini coefficient is associated with reduced height and weight, particularly among males, and decreases the sexual dimorphism in these anthropometric measures.

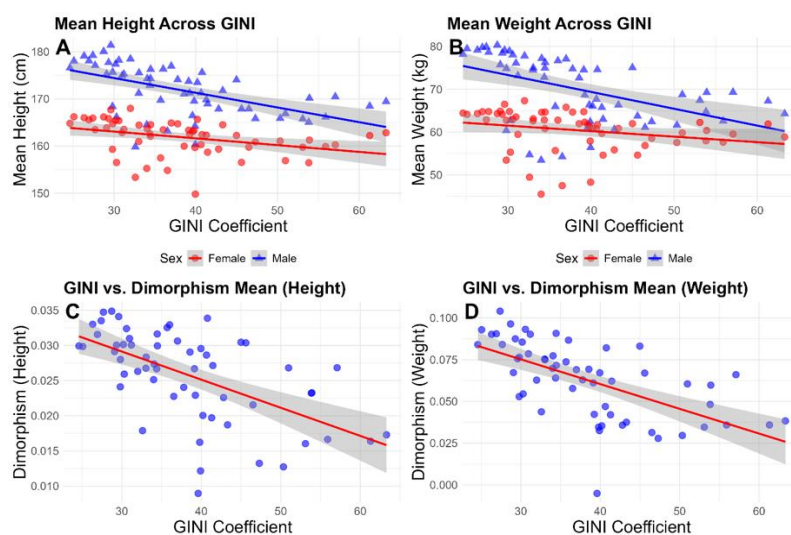

*Figure S04.* Panels A and B, respectively, show the relationships between Gini and mean height and weight for males (blue) and females (red). Panels C and D, respectively, show that dimorphism in height and weight increases with Gini. Linear fits are presented with 95% confidence intervals.

## Section 2. Kuh et al.'s analyses

### UK results excluding outlier standard deviations (SD)

The original SDs are plotted in Figure S05 (top part), and two of them are clearly different than the adjacent values and the trend lines. Critically, excluding these cohorts revealed the same pattern. The sex by cohort interaction was highly significant (p<.0002); women's SD did not change across cohorts ($B$=-.04, $p$=.264) and men's declined ($B$=-.23, $p$<.0005):

### Proportional Effects for Great Britian

Taking the height for the 1905 cohort as a baseline, we calculated proportional change across cohorts. The results confirmed the main analyses using cm. There is a 4.2% increase in height for men (from the earliest to latest cohorts), as compared to 1.9% for women (Figure S05, bottom). A regression confirmed a larger gain across the 10 cohorts for males ($b$=.40) than females ($b$=.16); $p$<.0001 for the interaction. The difference is close to the two-fold difference in gain found for cm.

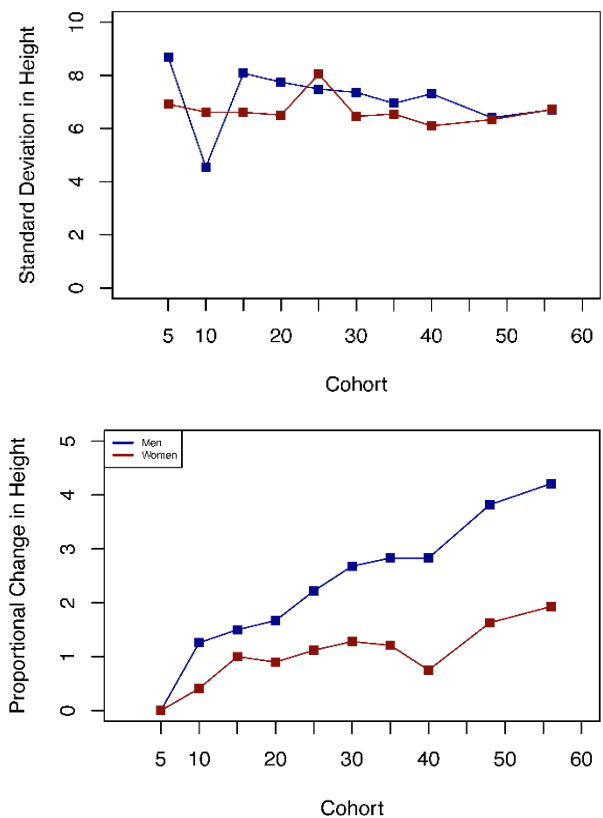

*Figure S05.* Standard deviations (top) and proportional change (bottom) in height (percent change, y-axis) across British cohorts.

**Section 3. Wikipedia analyses**

**Wiki data**

All sources provided on the Wikipedia page were checked for their veracity. Subsequently, the following four countries (representing 2.6% of the data) were not included in analyses: Ecuador, Israel, Kenya and Russia. Analysis of the Average Human Height by Country data confirmed the WHO data patterns. We compared the two beta coefficients using the classical method outlined by Cohen et al. in the 2003, $z = \frac{\beta_1 - \beta_2}{\sqrt{SE_{\beta_1}^2 + SE_{\beta_2}^2}}$.

This involved first calculating the difference between the beta coefficients from the two regression models. We then determined the standard error of this difference to account for the uncertainty in each beta estimate. Finally, we computed a Z-score to assess whether the observed difference was statistically significant, thereby determining if there was a meaningful difference in the relationship between the independent variable and the outcome across the models. The regression line for males explained almost twice the variance in height as did the regression line for females, and the former is almost twice as steep as the latter, males: $b$=25.58, $F(1, 126)$=98.59, $p$<.001, $R^2$=.439; females: $b$=13.73, $F(1, 145)$=37.49, $p$<.001, $R^2$=.205. A statistical comparison of the two betas indicates they were significantly different, $z$=3.47, $p$<.001.